# CATEGORIZATION OF STRINGED INSTRUMENTS WITH MULTIFRACTAL DETRENDED FLUCTUATION ANALYSIS

Archi Banerjee*, Shankha Sanyal, Tarit Guhathakurata, Ranjan Sengupta and Dipak Ghosh

Sir C.V. Raman Centre for Physics and Music
Jadavpur University, Kolkata: 700032

*archibanerjee7@gmail.com

* Corresponding Author

## ABSTRACT

*Categorization is crucial for content description in archiving of music signals. On many occasions, human brain fails to classify the instruments properly just by listening to their sounds which is evident from the human response data collected during our experiment. Some previous attempts to categorize several musical instruments using various linear analysis methods required a number of parameters to be determined. In this work, we attempted to categorize a number of string instruments according to their mode of playing using latest-state-of-the-art robust non-linear methods. For this, 30 second sound signals of 26 different string instruments from all over the world were analyzed with the help of non linear multifractal analysis (MFDFA) technique. The spectral width obtained from the MFDFA method gives an estimate of the complexity of the signal. From the variation of spectral width, we observed distinct clustering among the string instruments according to their mode of playing. Also there is an indication that similarity in the structural configuration of the instruments is playing a major role in the clustering of their spectral width. The observations and implications are discussed in detail.*

## INTRODUCTION

Classification is one of the processes involved in audio content description. Audio signals can be classified according to miscellaneous criteria viz. speech, music, sound effects (or noises). Usually, music streams are broadly classified according to genre, player, mood, or instrumentation [1]. Some of the most important instruments in the history of music have been stringed instruments, which range from early to modern forms of the violin and the guitar, through to contemporary experiments with amplification and electric or digital recording. Stringed instruments are often said to belong to different categories based on their timbre or how they produce sound. The string may be struck, plucked, rubbed (bowed), or, occasionally, blown (by the wind); in each case the effect is to displace the string from its normal position of rest and to cause it to vibrate in complex patterns. All these factors characterize the "timbre" of a stringed instrument. In a sense, timbre is everything that lets us distinguish one instrument from another. For example, a trumpet has a brash timbre compared with the smoother timbre of a saxophone. A violin and viola have very similar timbres (which is the reason we might struggle to distinguish them by ear, but can be distinguished through robust analysis). The timbre of a string instrument is completely different from a keyed or wind instrument. Timbre is a combination of many factors. It *includes tone* but also aspects like how *suddenly or smoothly* the notes start and end, the number and strength of *harmonics* in the sound, and how the sound *varies* over time. No system of classification can adequately categorize the interactions of natural material, craftsmanship, and exuberant imagination that produced an endless variety of stringed instruments. We envisaged developing a scientific technique involving only one or two parameters with the help of which we can automatically ascertain that a particular stringed instrument belongs to a certain family of stringed instruments. This automated technique also clearly characterizes groups of instruments which seem to hear similar to human ear but is distinguishable with the method developed.

Mandelbrot [2] demonstrated how nature contains structures (e.g., mountains, coastlines, the structures of plants), which could be described by fractals and suggested that fractal theory could be used in order to understand the harmony of nature. Fractals can also be found in other natural processes described by time-series measurements (i.e., noises, pitch and loudness variations in music, demographic data and others). Analysis of musical structure has revealed evidence of both fractal aspects and self-similarity properties in instrument tones and music genres. Voss and Clark [3] investigated aspects in music and speech by estimating the power spectra for slowly varying quantities, such as loudness and frequency. The fractal and multifractal aspects of different genres of music were analyzed by Bigrelle and Iost [4], where it was proposed that the use of fractal dimension measurements could benefit the discrimination of musical genres. Su and Wu [5] applied Hurst exponent and Fourier analysis in sequences of musical notes and noted that music shares similar fractal properties with the fractional Brownian motion. But, music signals may exhibit self similarity in different scales which may not be described by single scaling exponent as is found in Detrended Fluctuation Analysis (DFA) technique [6]. Such a system is better characterized as 'multifractal'. A multifractal can be loosely thought of as an interwoven set constructed from sub-sets with different local fractal dimensions. Real world systems are mostly multifractal in nature. Music too, has non-uniform property in its movement. Since the multifractal technique analyzes the signal in different scales, it is able to decipher much accurately the amount of self similarity present in a signal. The spectrum in multifractal detrended fluctuation analysis (MFDFA) is the measure of complexity or self-similarity present in the signal [7]. This method is very useful in the analysis of various non-stationary time series and it also gives information regarding the multifractal scaling behaviour of non-stationary signals. We used this technique to characterize signals taken from various stringed instruments and the spectral width obtained from each case has been used as a parameter to classify each instrument. A human response data was also taken for these instruments which were used for comparison purpose with the computational algorithm.

## EXPERIMENTAL DETAILS

In this work, we collected recorded clips of 26 stringed instruments (belonging to different categories according to their mode of playing) from all over the world. Each of the instruments was played in their own/traditional style and segments of 30 seconds were cut from each clip for analysis. Among the 26 string instruments used in the experiment, 21 were plucked string instruments, 3 were bowed while 2 were struck string instruments. Amplitudes of all the clips were normalized to 0 dB. All the signals were digitized at the rate of 44100 samples/sec 16 bit format. Multifractal spectral width (W) was calculated for all the 30s clips collected from 26 different stringed instruments using the MFDFA technique. Then, the 26 instrumental signals were played (without revealing their mode of playing to the listeners) in a random order using a sound system with high S/N ratio and a human response data from of 30 naïve listeners were taken. The subjects were asked to mark the clips to which category of stringed instruments they felt the clips fell – plucked, struck or bowed. A confusion matrix was created from the diverse response that was obtained from the listening test data.

## METHOD OF ANALYSIS
### Method of multifractal analysis of sound signals

The time series data obtained from the sound signals are analyzed using MATLAB [8] and for each step an equivalent mathematical representation is given which is taken from the prescription of Kantelhardt et al [7].

The complete procedure is divided into the following steps:

*Step 1:* Converting the noise like structure of the signal into a random walk like signal. It can be represented as:

$$Y(i) = \sum (x_k - \bar{x}) \quad (1)$$

Where $\bar{x}$ is the mean value of the signal.

*Step 2:* The whole length of the signal is divided into Ns number of segments consisting of certain no. of samples. For s as sample size and N the total length of the signal the segments are

$$Ns = \text{int}\left(\frac{N}{s}\right) \quad (2)$$

*Step 3:* The local RMS variation for any sample size *s* is the function *F(s,v)*. This function can be written as

follows:

$$F^2(s,v) = \frac{1}{s}\sum_{i=1}^{s}\{Y[(v-1)s+i] - y_v(i)\}^2$$

*Step 4:* The q-order overall RMS variation for various scale sizes can be obtained by the use of following equation

$$F_q(s) = \left\{\frac{1}{Ns}\sum_{v=1}^{Ns}[F^2(s,v)]^{\frac{q}{2}}\right\}^{\left(\frac{1}{q}\right)} \quad (3)$$

*Step 5:* The scaling behaviour of the fluctuation function is obtained by drawing the log-log plot of $F_q(s)$ vs. s for each value of q.

$$F_q(s) \sim s^{h(q)} \quad (4)$$

The h(q) is called the generalized Hurst exponent. The Hurst exponent is measure of self-similarity and correlation properties of time series produced by fractal. The presence or absence of long range correlation can be determined using Hurst exponent. A monofractal time series is characterized by unique h(q) for all values of q.

The generalized Hurst exponent h(q) of MFDFA is related to the classical scaling exponent τ(q) by the relation

$$\tau(q) = qh(q) - 1 \quad (5)$$

A monofractal series with long range correlation is characterized by linearly dependent q order exponent τ(q) with a single Hurst exponent H. Multifractal signal on the other hand, possess multiple Hurst exponent and in this case, τ(q) depends non-linearly on q [9].

The singularity spectrum f(α) is related to h(q) by

$$\alpha = h(q) + qh'(q)$$
$$f(\alpha) = q[\alpha - h(q)] + 1$$

Where α denoting the singularity strength and *f(α)*, the dimension of subset series that is characterized by α. The width of the multifractal spectrum essentially denotes the range of exponents. The spectra can be characterized quantitatively by fitting a quadratic function with the help of least square method [9] in the neighbourhood of maximum $\alpha_0$,

$$f(\alpha) = A(\alpha - \alpha_0)^2 + B(\alpha - \alpha_0) + C \quad (6)$$

Here C is an additive constant C = f($\alpha_0$) = 1 and B is a measure of asymmetry of the spectrum. So obviously it is zero for a perfectly symmetric spectrum. We can obtain the width of the spectrum very easily by extrapolating the fitted quadratic curve to zero.

Width W is defined as,

$$W = \alpha_1 - \alpha_2 \quad (7)$$

with $f(\alpha_1) = f(\alpha_2) = 0$

The width of the spectrum gives a measure of the multifractality of the spectrum. Greater is the value of the width W greater will be the multifractality of the spectrum. For a monofractal time series, the width will be zero as h(q) is independent of q. The spectral width has been considered as a parameter to evaluate how a group of string instruments vary in their pattern of playing from another

## RESULTS AND DISCUSSION

The multifractal spectral width (W) is said to be the manifestation of complexity of that particular signal. If the spectral width of a number of instruments is found to be comparable, those instruments are said to form a cluster. Thus, we have found a number of clusters of spectral widths corresponding to particular families of string instruments. Table 1 gives the multifractal spectral widths and shuffled widths for all the 26 instrumental clips along with their mode of playing. Fig. 1 graphically plots the different clusters that are formed by the instruments in regard to their multifractal spectral width.

**Table 1:** Clustering of spectral widths in different groups for all 26 instruments

| Type of Instrument | Group | INSTRUMENT | SPECTRAL WIDTH |
|---|---|---|---|
| PLUCKED | Group 1 | Banjo | 0.672 |
| | | Cittern | 0.630 |
| | | Dotara | 0.643 |
| | | Ektara | 0.723 |
| | | Harp | 0.686 |
| | | Hawaian Guitar | 0.632 |
| | | Kora | 0.626 |
| | | Mohan Veena | 0.717 |
| | | Spanish Guitar | 0.749 |
| | | English Guitar | 0.716 |
| | | Portuguese Guitar | 0.584 |
| | | Harpsichord | 0.675 |
| | | Sarod | 0.702 |
| | | Murchang | 0.715 |
| | Group 2 | Sitar | 0.428 |
| | | Veena | 0.483 |
| | | Rudra Veena | 0.387 |
| | | Surbahar | 0.495 |
| | | Tanpura | 0.506 |
| | Group 3 | Mandolin | 0.824 |
| | | Mandriola | 0.812 |
| BOWED | Group 4 | Esraj | 0.803 |
| | | Rawanhata | 0.896 |
| | | Sarengi | 0.912 |
| STRUCK | Group 5 | Piano | 0.531 |
| | | Santoor | 0.515 |

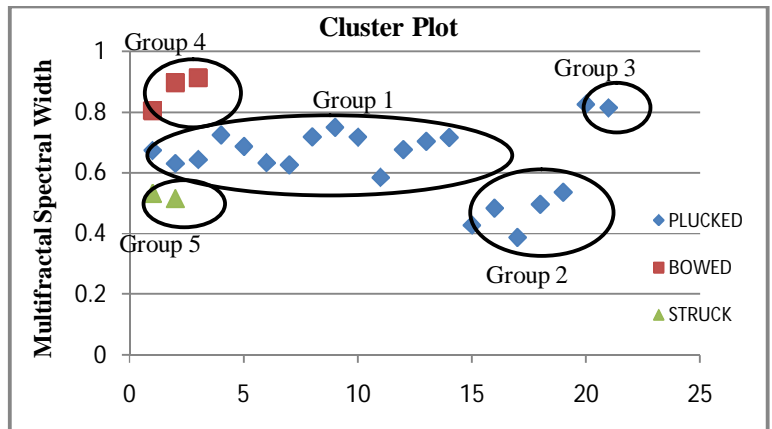

**Fig 1:** Formation of clusters in spectral width values among different string instruments

1) From Table 1 and Fig 1, it is evident that the chosen string instruments can be divided into distinct groups according to the three different modes of playing (plucking, striking or bowing the strings) from their spectral width variation. The spectral widths of the struck string and bowed string category clustered around 0.50-0.55 (Group 5) and 0.80-0.90 (Group 4) respectively while there were at least 3 clusters even within the plucked string category whose spectral widths varied around 0.55-0.75 for Group 1, 0.35-0.50 for Group 2 and 0.80-0.85 from Group 3. Formation of the 3 groups among the plucked string category may be caused due to their structural difference as table 1 shows that the instruments with *'sitar'*-like structure are forming a cluster in terms of spectral width. Also, within Group 1 we found that *'guitar'*-like instruments falling within the same cluster along with some other instruments like *sarod, ektar, dotar* etc. In Group 3, another two similar looking instruments *mandolin* and *mandriola* form a cluster.
2) The presence of instruments like *sarod, ektar, dotar* and *banjo, harp* etc. along with the *'guitar'* group in the same cluster is quite an interesting observation. These categories of instruments are structurally very much different from one another which qualitatively demands that they should fall in different categories. But, their clustering in the same group indicates the presence of some inherent similarity in the characteristic of complex sound patterns produced by them. This provides scientific proof regarding the fact that instruments which are very much different in their structure may have similar timbral quality.

On the other hand, the human response data may serve as the working baseline for us as how the perception of different categories of stringed instruments varied to the respondents. The following confusion matrix in Table 2 can give us an idea about the instruments which were perceived correctly by most of the listeners according to their way of playing and those which created more confusion in listeners' minds.

**Table 2:** Confusion matrix calculated from human response data obtained from listening test

|  | Plucked (%) | Struck (%) | Bowed (%) |
|---|---|---|---|
| **Plucked** | 73.14 | 23.14 | 3.71 |
| **Struck** | 22 | 78 | 0 |
| **Bow** | 4 | 1 | 95 |

1) From Table 2 it is evident that 95% of the subjects perceived bowed instruments correctly whereas only 1% and 4% of the subjects confused them with struck and plucked string instruments respectively. On the other hand 78% of the subjects identified struck string instruments correctly; rest 22% confused them with plucked category while no one perceived them as bowed string instruments. But in case of plucked string instruments the rate of confusion was higher among the said 3 categories. 73.14% subjects could identify the plucked string instruments correctly, 23.14% and 3.14% of the subjects confused them as struck and bowed string instruments respectively.
2) From the detailed analysis of listening test human response data we found –
   i. In Plucked string category: Most distinguishable – Cittern; Most confusing – Harp
   ii. In Struck string category: Most distinguishable – Piano; Most confusing – Santoor
   iii. In Bowed string category: Most distinguishable – Esraj, Rawanhata; Most confusing – Sarengi

Comparing human response data and spectral width data obtained applying MFDFA technique we can say that those instruments which were easily confused by the listeners during the listening test can be easily identified from their spectral width to fall in the proper category according to their way of playing.

## CONCLUSION
In the human response data, when the subjects were listening to the chosen instrumental clips, there was sufficiently large amount of confusion in identifying the plucked and struck string instruments separately whereas the overlap in perception between bowed string instruments with other 2 categories is much lesser. But this confusion could be eliminated from the multifractal analysis of the clips. As a part of this analysis, multifractal spectral widths were determined for all the clips. The instruments, whose complexity values are in close proximity, are said to form a cluster. We hypothesize that the instruments which fall in a particular cluster have similar timbral features, which are significantly different from other stringed instruments. Thus, separate clusters were formed for different modes of playing as well as for different groups of structural resemblances among those instruments. This technique of categorising string instruments using robust non-linear MFDFA method has not been reported in literature till now. The inherent difficulty associated with this experiment is the absence of large number of struck string instruments throughout the world. Also many local bowed string instruments were not available for analysis during our experiment. A larger data base including more number of clips in each category needs to be analysed using this technique to obtain a more conclusive results with greater number of clusters or more weightage of clips within the same cluster. Various wind and percussion instruments should be analysed also for identifying the clusters within these two categories as well as to study the overlaps (if any) between string, wind and percussion instruments.

To conclude, we have developed a novel technique using an automated algorithm, which can be further applied to different categories of musical instruments for distinguishing them on the basis of their timbral features, which is not possible by human ear.